\documentstyle[eqsecnum,preprint,aps,epsfig]{revtex}
\tightenlines

\newcommand{\beq}{\begin{equation}}
\newcommand{\eeq}{\end{equation}}
\newcommand{\beqa}{\begin{eqnarray}}
\newcommand{\eeqa}{\end{eqnarray}}
\def\nn{\nonumber}
\def\l({\left(}
\def\r){\right)}

\title{Brane collisions in anti-de Sitter space.}
\author{Andrey Neronov
\footnote{e-mail address: neronov@theorie.physik.uni-muenchen.de}}
\address{Theoretische Physik, Universit\"at M\"unchen, \\
Theresienstr., 37, 
80333, Munich, Germany} 
\begin{document}

\maketitle
\draft

\narrowtext

\begin{abstract}
From the requirement of continuous matching of bulk metric around the point 
of brane collision we derive a conservation law  for collisions of p-branes in 
(p+2)-dimensional space-time. 
This conservation law relates energy densities on the 
branes before and after the collision. Using this conservation law we 
are able to calculate the amount of matter produced in the collision 
of orbifold-fixed brane with a bulk brane in the ``ekpyrotic/pyrotechnic 
type''  models of brane cosmologies.
\end{abstract}


\section{Introduction.}
In several recently proposed models of brane cosmology 
\cite{ekpirots,pyro,quevedo,bucher} the creation of four-dimensional 
expanding Friedman-Robertson-Walker (FRW) universe is associated with 
collisions of branes in higher-dimensional space-time. For example, in 
ekpyrotic/pyrotechnic scenario \cite{ekpirots,pyro} 
the visible universe is represented by a 
negative/positive tension 3-brane sitting on a fixed point of $Z_2$ orbifold
in five-dimensional space-time. The orbifold 
brane collides inelastically with a brane moving through the bulk. At the 
moment of collision some amount of relativistic matter (radiation) is produced 
on the visible brane at the cost of kinetic energy of the bulk brane. Initially
flat Minkowsky universe residing on the visible brane starts to expand. Thus, 
the initial conditions for the hot FRW universe are generated during the 
brane collision. An immediate problem to be addressed in these scenarios is 
what is the energy density of radiation (or, equivalently, what is the 
initial temperature of the universe) produced on the visible brane? 
Another question is  whether 
the visible brane expands or collapses right after the collision? Indeed, 
naively, if one deposits homogeneous energy distribution into initially 
static universe, it will start to collapse.  

In what follows we find a conservation law for the brane collisions which 
provides answers to the above questions. This conservation law is a direct 
generalization of a conservation law found by Dray and 't Hooft \cite{dray}
for the case of collisions of light-like shells in four-dimensional 
Schwarzschild space-time.

\section{Conservation law in brane collisions.}

\vspace{5mm}
\centerline{\epsfig{file=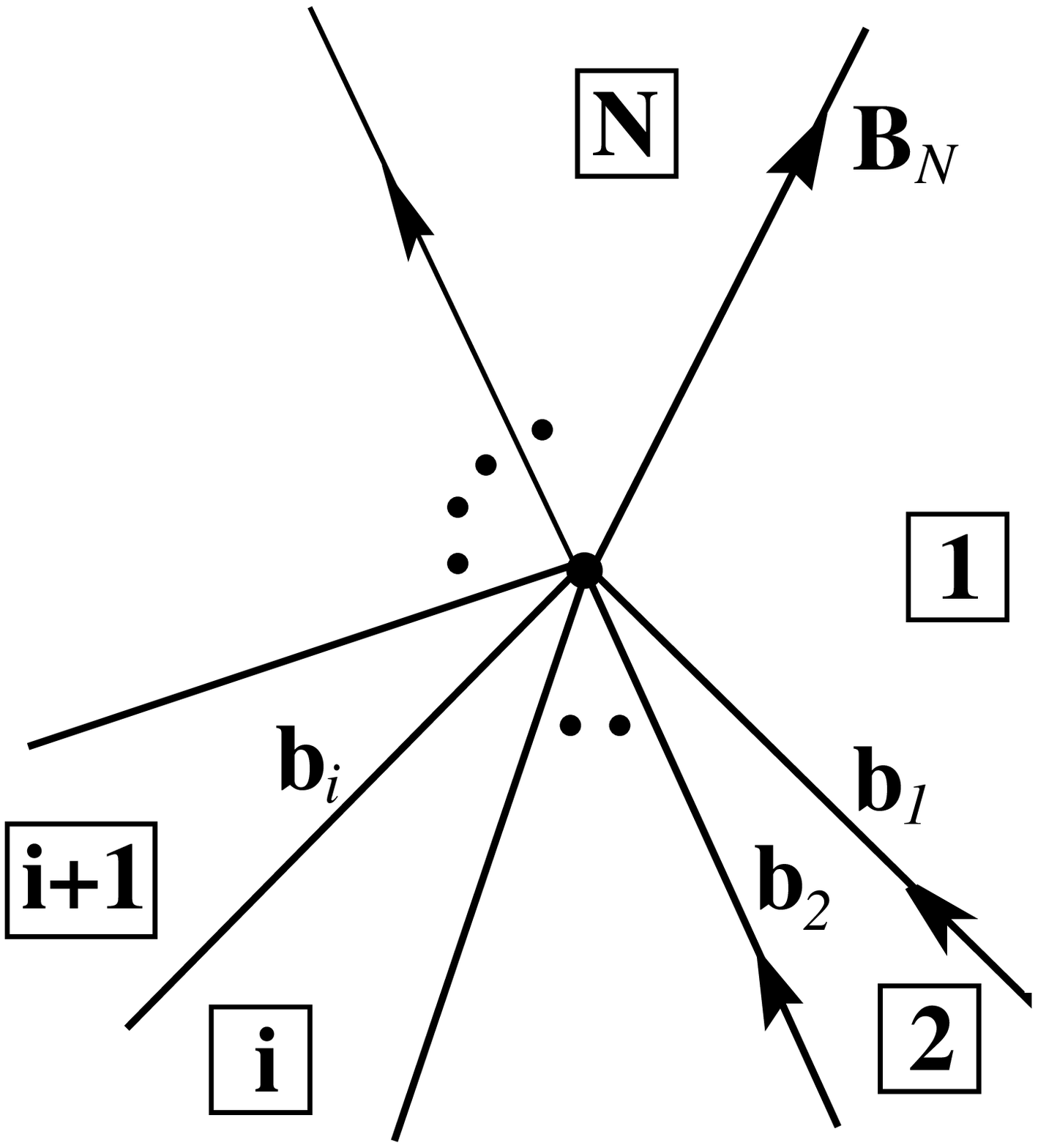,width=6cm}}
{\footnotesize\textbf{Figure 1:} Colliding branes.}
\vspace{5mm}

Let us consider a collision of $k$ in-branes ${\bf b}_1,..,{\bf b}_k$ 
in which $N-k$ out-branes ${\bf B}_{k+1}, .., {\bf B}_N$ are formed, as it is 
shown on Fig. 1.  The $N$ branes 
divide the space-time onto $N$ disjoint regions {\bf 1,..,N}. 
Suppose that the whole 
space-time  possesses a three-dimensional Eucledian symmetry $E(3)$
so that  the metric in each region {\bf i}
can be written in isotropic coordinates $(U_i, V_i)$ as
\beq
\label{region-i}
ds^2=-F_i(U_i, V_i)dU_idV_i+H_i(U_i, V_i)\l(dx_1^2+dx_2^2+dx_3^2\r)
\eeq
where $F_i, H_i$ are some functions of $(U_i, V_i)$ and 
$(x_1, x_2, x_3)$ are coordinates of three-dimensional Eucledian space.
We can always rescale the coordinates $x_k$ in such a way that $H_i=1$ at 
the collision point.

If we want the space-time metric in the neighborhood of the collision point to
be well-defined, we have to suppose that there exist isotropic coordinates 
$(u, v)$ in which the space-time metric has the form  
\beq
\label{c0}
ds^2=-f(u,v)dudv+h(u,v)\l(dx_1^2+dx_2^2+dx_3^2\r)
\eeq
and the functions $f(u, v), h(u, v)$ are continuous at the collision point 
$(u_c, v_c)$.

In each region {\bf i} the coordinates $(U_i, V_i)$ can be expressed through 
$(u, v)$
\beq
\label{u,v}
U_i=U_i(u), \ \ \ V_i=V_i(v)
\eeq
Comparing the  metrics (\ref{region-i}) and (\ref{c0}) we find 
that the functions $f, h$ are 
related to the functions $F_i, H_i$ as 
\beqa
\label{match-f}
f(u, v)&=&F_i\l(U_i(u), V_i(v)\r)U_i'(u)V_i'(v)\nn\\
h(u, v)&=&H_i\l(U_i(u), V_i(v)\r)
\eeqa
where prime denotes the derivatives of functions (\ref{u,v}). In particular, 
if we restrict (\ref{match-f}) to the brane ${\bf b}_i$ we find
\beqa
\label{match-F}
\left.F_i(U_i, V_i)U_i'V_i'\right|_{{\bf b}_i}&=&
\left.F_{i+1}(U_{i+1}, V_{i+1})U_{i+1}'V_{i+1}'\right|_{{\bf b}_i}\\
\label{match-H}
\left.H_i(U_i, V_i)\right|_{{\bf b}_i}&=&
\left.H_{i+1}(U_{i+1}, V_{i+1})\right|_{{\bf b}_i}
\eeqa
The brane ${\bf b}_i$ which is the boundary between regions 
{\bf i} and {\bf i+1} is a surface defined by equation
\beq
\Sigma_i^+\l(U_i, V_i\r)=0
\eeq
in the coordinates of the region {\bf i} or 
\beq
\Sigma_{i+1}^-\l(U_{i+1}, V_{i+1}\r)=0
\eeq
in the coordinates of the region {\bf i+1}. If we take $u$ as one of the 
coordinates along the brane ${\bf b}_i$ and take  $u$ derivative of 
(\ref{match-H}) along ${\bf b}_i$ we find
\beq
\label{someth}
\l(\partial_{U_i}H_i-\partial_{V_i} H_i\frac{\partial_{U_i}\Sigma_i^+}{
\partial_{V_i}\Sigma_i^+}\r)U_i'=
\l(\partial_{U_{i+1}}H_{i+1}-
\partial_{V_{i+1}} H_{i+1}\frac{\partial_{U_{i+1}}\Sigma_{i+1}^-}{
\partial_{V_{i+1}}\Sigma_{i+1}^-}\r)U_{i+1}'
\eeq  
Taking (\ref{someth}) and (\ref{match-F})
at the collision point $(u_c, v_c)$ we find relations between 
$U_i', V_i'$ and $U_{i+1}', V_{i+1}'$ 
\beq
\l(\begin{array}{c}
U_{i+1}'\\
V_{i+1}'
\end{array}\r)={\cal C}_{i(i+1)}
\l(\begin{array}{c}
U_{i}'\\
V_{i}'
\end{array}\r)
\eeq
where ${\cal C}_{i(i+1)}$ is a numerical $2\times 2$ matrix. The transition 
matrices ${\cal C}_{12},..,{\cal C}_{(N-1)N},{\cal C}_{N1}$ can be calculated 
for each pair of adjacent regions. An obvious consistency condition 
which must be satisfied 
if there exist coordinates $(u, v)$ in the neighborhood of collision
point such that the metric (\ref{c0}) is continuous at 
$(u_c, v_c)$ is
\beq
\label{consist}
{\cal C}_{12}...{\cal C}_{(N-1) N}{\cal C}_{N 1}=I
\eeq
where $I$ is the identity matrix. In what follows we show that 
this consistency condition provides us with a nontrivial 
conservation law, which relates the physical 
parameters of the in-branes ${\bf b}_1,..,{\bf b}_k$ to the 
parameters of the out-branes ${\bf B}_{(k+1)},..,{\bf B}_N$.

We consider the bulk metric of the form
\beq
\label{general}
ds^2=-F(R)dT^2+\frac{dR^2}{F(R)}+R^2\l(\sum_{k=1}^3 dx_k^2\r)
\eeq
where $F(R)$ is a given function. Substituting (\ref{general}) in the 
bulk Einstein  equations with a negative cosmological constant 
$-\Lambda$ we find a general solution
\beq
F(R)=\Lambda R^2-\frac{M}{R^2}
\eeq
where $M$ is an integration constant. 
Let us write the condition (\ref{someth}) for this metric in more detailed 
form. In order to go to the isotropic coordinates $(U, V)$ in the metric 
(\ref{general}) we make a coordinate change
\beq
\label{z}
dz=-\frac{dR}{F(R)}
\eeq
so that the metric becomes
\beq
ds^2=F(R)\l(-dT^2+dz^2\r)+R^2\l(\sum_{k=1}^3 dx_k^2\r)
\eeq
The coordinates $(U, V)$ are then
\beqa
U&=&T-\sigma z(R)\nn\\
V&=&T+\sigma z(R)
\eeqa
where $\sigma=+1$ if $z$ grows to the right and $\sigma=-1$ is $z$ grows to 
the left (if a space-time region {\bf i}  is a $Z_2$ image of a 
space-time region {\bf j} with $\sigma_j=+1$ then $\sigma_i=-1$ 
and vise versa).

The trajectories of branes are given by the functions
\beq
\label{traj}
\Sigma=R-\bar R(T)=0
\eeq
where $\bar R(T)$ is found from the equations of motion of the brane.
Substituting (\ref{traj}) into the l.h.s. or r.h.s. of the matching 
condition (\ref{someth}) we find
\beq
\label{s1}
\l(\partial_{U}H-\partial_{V} H\frac{\partial_{U}\Sigma}{
\partial_V\Sigma}\r)=\frac{2RF\bar R'}{(F+\sigma \bar R')}
\eeq
Introducing the proper time along the brane through 
the relation
\beq
d\tau^2=\l(F-\frac{\left.\bar R'\right.^2}{F}\r)dT^2
\eeq
we can rewrite (\ref{s1}) as
\beq
\l(\partial_{U}H-\partial_{V} H\frac{\partial_{U}\Sigma}{
\partial_{V}\Sigma}\r)=\frac{2RF\dot{\bar R}}{\l(\dot{\bar R}^2+
F\r)^{1/2}+\sigma\dot{\bar R}}
\eeq
where dot denotes the proper time derivative.
Taking the values of the last 
expression for regions {\bf i} and {\bf i+1} we find form
(\ref{someth}) 
\beq
\label{maa}
U_{i+1}'=\left[
\frac{F_i\l(\l(\dot{\bar R}^2+F_{i+1}\r)^{1/2}+\sigma \dot{\bar R}\r)}
{F_{i+1}\l(\l(\dot{\bar R}^2+F_i\r)^{1/2}+\sigma \dot{\bar R}\r)}\right]U_i'
\eeq
if $\sigma_i=\sigma_{i+1}=\sigma$ or  
\beq
\label{maaa}
U_{i+1}'=\left[\frac{\l(\dot{\bar R}^2+F_i\r)^{1/2}-\sigma \dot{\bar R}}
{\l(\dot{\bar R}^2+F_i\r)^{1/2}+\sigma \dot{\bar R}}\right]U_i'
\eeq
if $\sigma_i=-\sigma_{i+1}=\sigma$.

\vspace{5mm}
\centerline{\epsfig{file=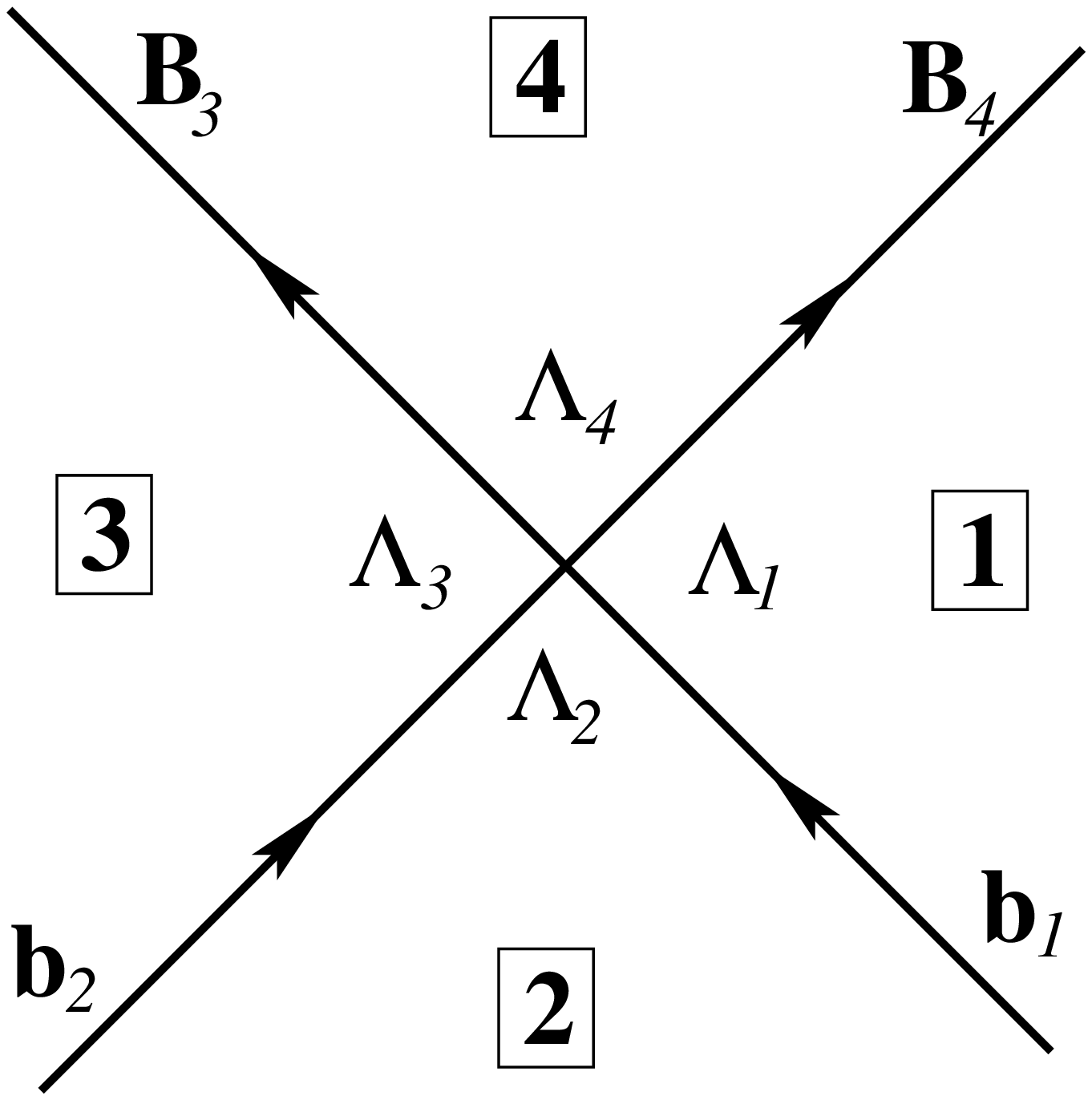,width=5cm}}
{\footnotesize\textbf{Figure 2:} Collision of light-like branes.}
\vspace{5mm}

A simple example of brane collisions is presented on Figures 2. 
The light-like branes ${\bf b}_1, {\bf b}_2, {\bf B}_3, {\bf B}_4$ are 
boundaries between the regions  
{\bf 1, 2, 3} and {\bf 4} which are all patches of anti-deSitter space-time 
with different 
cosmological constants $\Lambda_1, \Lambda_2, \Lambda_3, \Lambda_4$. 
In this case the consistency condition (\ref{consist}) gives (see 
Appendix for details of calculation) 
\beq
\Lambda_1\Lambda_3=\Lambda_2\Lambda_4
\eeq

\section{Collision of a bulk brane with an orbifold-fixed brane.}

\vspace{5mm}
\centerline{\epsfig{file=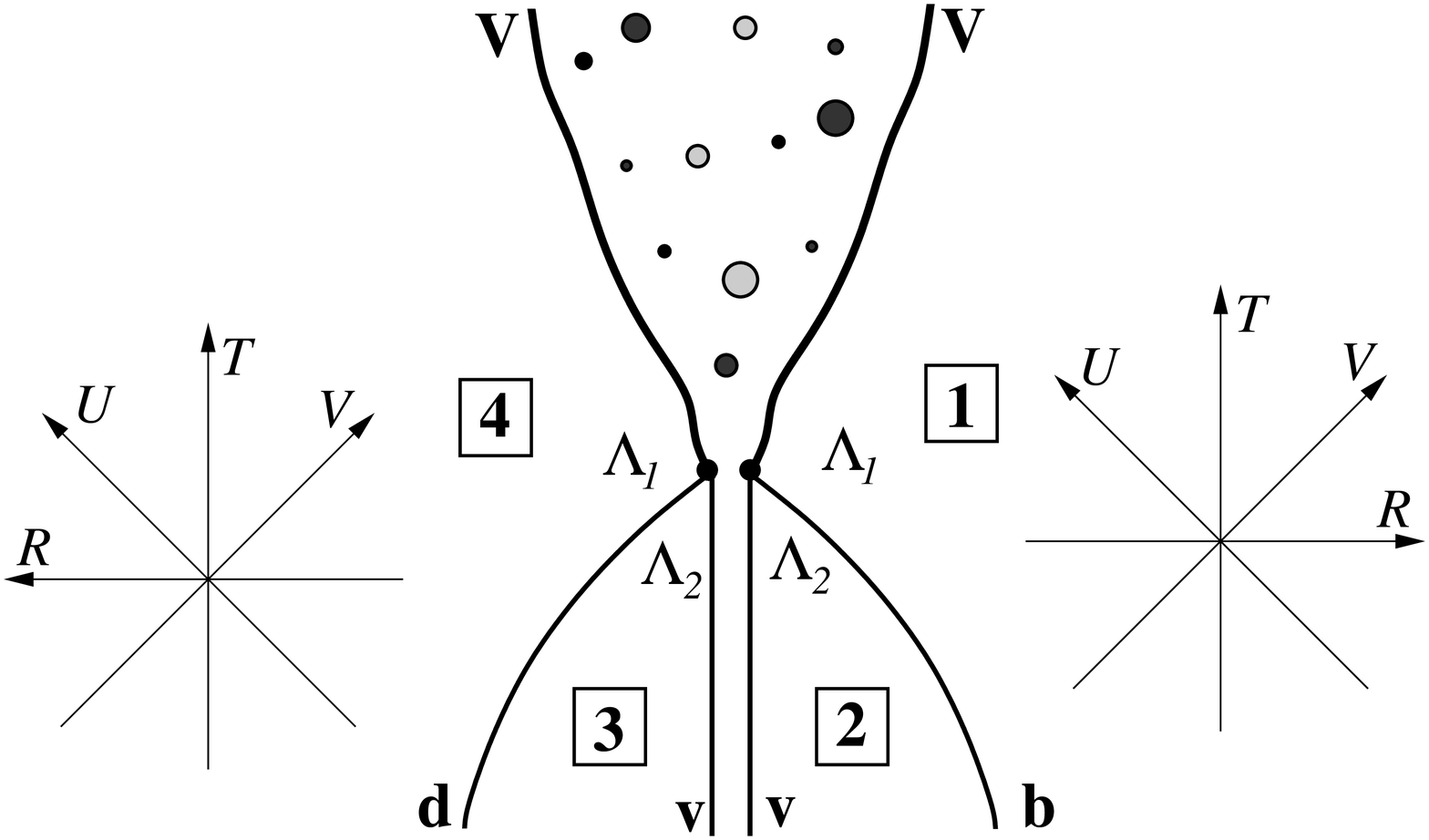,width=11cm}}
{\footnotesize\textbf{Figure 3:} Collision of a bulk brane {\bf b} with an 
orbifold brane {\bf v}.}
\vspace{5mm}

We proceed with a more complicated example in which a $Z_2$-orbifold 
brane {\bf v}  is hit inelastically by a bulk brane 
{\bf b}, as it is shown on 
Fig. 3. It is convenient to unfold $Z_2$ orbifold so that the 
space-time consists of four regions, the region {\bf 3} is a mirror copy 
of the region {\bf 2} and the region {\bf 4} is a mirror copy of the region 
{\bf 1}. The situation presented on Fig. 3 
was discussed recently in the ``ekpyrotic'' and ``pyrotechnic''  scenarios
of brane cosmology.

We can make different assumptions about the bulk metric in regions 
{\bf 1} and {\bf 2}. For example, we can suppose that the cosmological 
constants in these regions are different, $\Lambda_1$ and $\Lambda_2$ 
respectively. Otherwise we can suppose that the cosmological 
constants are the same, $\Lambda_1=\Lambda_2=\Lambda$, but the metric in 
region {\bf 1} is different from anti-deSitter
\beq
\label{f1}
F_1=\Lambda R^2-\frac{M_1}{R^2}
\eeq
we will consider both possibilities.

In order to write down the matching conditions (\ref{maa}) or (\ref{maaa}) 
between different regions we need to specify the value of parameter 
$\sigma$ in each region. If we take $\sigma=+1$ in the regions {\bf 1}
and {\bf 2} this will automatically imply the $\sigma=-1$ in the regions 
{\bf 4} and {\bf 3} since these regions are mirror copies of {\bf 1, 2}.
In this case the coordinate $z$ (\ref{z}) grows to the right in the region 
{\bf 1} while the coordinate $R$
\beq
R=\frac{1}{\Lambda z}
\eeq
grows to the left. One can check that the tensions 
of the visible branes {\bf v, V} are positive in this case. 
If we take $\sigma=-1$ in the regions {\bf 1, 2} the visible branes have 
negative tension. The latter situation corresponds to the one 
considered in the original ekpyrotic scenario \cite{ekpirots} while 
the former has been analyzed in pyrotechnic model \cite{pyro}.

The trajectory of the bulk brane {\bf b} is $\bar R_b(\tau)$. It is convenient 
to introduce the expansion rate 
\beq
H_b=\frac{\dot{\bar R}_b}{\bar R_b}
\eeq
where dot denotes the proper time derivative.
The matching condition (\ref{maa}) between the regions {\bf 1} and {\bf 2} 
reads
\beq
\label{12}
U_2'=\left[\frac{F_1(\sqrt{H_b^2+F_2/R^2}+\sigma H_b)}
{F_2(\sqrt{H_b^2+F_1/R^2}+\sigma H_b)}\right]U_1'
\eeq

The trajectory of the visible brane {\bf V} after the collision 
is $\hat R_V(\tau)$ and its expansion rate is $H_V$. 
Since the region {\bf 4} is a mirror copy of the region
{\bf 1} we use the condition (\ref{maaa}) for matching the regions 
{\bf 1} and {\bf 4}
\beq 
\label{14}
U_4'=\left[\frac{\sqrt{H_V^2+F_1/R^2}-\sigma H_V}
{\sqrt{H_V^2+F_1/R^2}+\sigma H_V}\right]U_1'
\eeq

The trajectory of the visible brane {\bf v} before the collision is 
$\bar R_v(\tau)=const$ and its expansion rate is, correspondingly, 
$H_v=0$. The matching condition between regions {\bf 2} and {\bf 3} 
is obtained from (\ref{14}) by taking $H_V\rightarrow 0$
\beq
U_3'=U_2'
\eeq

Finally, the trajectory of the brane {\bf d} which is the mirror copy of 
the brane {\bf b} is $\bar R_v(\tau)$. The values of parameter 
$\sigma$ in the regions {\bf 2} and {\bf 3} are 
$\sigma_{2,3}=-\sigma_1=\sigma$. From the matching condition (\ref{maa}) we 
find
\beq
\label{34}
U_4'=\left[\frac{F_2(\sqrt{H_b^2+F_1/R^2}-\sigma H_b)}
{F_1(\sqrt{H_b^2+F_2/R^2}-\sigma H_b)}\right]U_3'
\eeq

Combining (\ref{12})--(\ref{34}) we find
\beq
\label{final}
\frac{
(\sqrt{H_V^2+F_1/R^2}+\sigma H_V)
(\sqrt{H_b^2+F_1/R^2}-\sigma H_b)
(\sqrt{H_b^2+F_2/R^2}+\sigma H_b)
}
{
(\sqrt{H_V^2+F_1/R^2}-\sigma H_V)
(\sqrt{H_b^2+F_2/R^2}-\sigma H_b)
(\sqrt{H_b^2+F_1/R^2}+\sigma H_b)
}=1
\eeq
In order to simplify the above expression, let us suppose that 
$H_b^2, H_V^2$ are small compared to the $F_1/R^2, F_2/R^2$. Then 
in the first approximation we get 
\beq
\label{appro}
H_V\approx H_b\l(1-\sqrt{\frac{F_1}{F_2}}\r)
\eeq  
The consistency condition (\ref{final}) or (\ref{appro}) can be treated as a 
sort of energy conservation law for the brane collision. Indeed, the 
expansion rate  $H$ is related to the energy 
density $\epsilon$ on the brane 
through the five-dimensional analog of Friedman equation \cite{neronov}
\beq
\label{berezin}
\sigma_{r}\sqrt{H^2+F_r(R)/R^2}-\sigma_{l}\sqrt{
H^2+F_l(R)/R^2}=\epsilon 
\eeq
where the indexes $l, r$ refer to the regions on the left and on the right 
from the brane. For the visible brane {\bf V} $\sigma_r=-\sigma_l=\sigma$ 
and $F_l=F_r=F_1$. The energy density $\epsilon$ consists of the brane tension
$\lambda_V$ and energy density of radiation $\epsilon_{rad}$ 
produced at the moment of collision. For $H^2_V\ll F_1/R^2$ and 
$M_1\ll \Lambda_1$ we find 
\beq
\label{eps}
\epsilon_{rad}\approx
2\sigma\sqrt{\Lambda_1}\l(
1+\frac{H_V^2}{2\Lambda_1}-\frac{M_1}{2\Lambda_1 R^4}\r)
-\lambda_V
\eeq 
If the tension of the visible brane is fine 
tuned to the  bulk cosmological constant
$\lambda_V=2\sigma\sqrt{\Lambda_1}$
then
\beq
\label{las}
\epsilon_{rad}\approx\sigma\l(
\frac{H_V^2}{\sqrt{\Lambda_1}}-\frac{M_1}{\sqrt{\Lambda_1}R^4}\r)
\eeq 

If we take the visible brane of positive tension, $\sigma=1$, as it is 
in pyrotechnic scenario \cite{pyro} and the bulk 
metric in the region {\bf 1} to be anti-deSitter, $M_1=0$, we find from 
(\ref{appro}) and (\ref{las}) that 
energy density of matter produced on the brane after the collision is
\beq
\label{result1}
\epsilon_{rad}\approx H_b^2\l(\frac{1}{\sqrt{\Lambda_1}}-\frac{1}{
\sqrt{\Lambda_2}}\r)
\eeq
And the initial expansion rate of the brane {\bf V} is
\beq
H_V\approx H_b\l(1-\sqrt{\frac{\Lambda_1}{\Lambda_2}}\r)
\eeq
Thus, if $\Lambda_2>\Lambda_1$ some positive energy density 
is produced in the universe residing on the visible brane
and the universe starts to expand (since the expansion rate of the bulk 
brane $H_b$ was positive at the moment of collision).

If we consider the visible brane of negative tension, $\sigma=-1$, as it is 
in original ekpyrotic model of \cite{ekpirots} and, for example choose 
$M_1\not= 0$ in the region {\bf 1} we find
\beq
\epsilon_{rad}\approx \frac{M_1}{\sqrt{\Lambda_1}}
\eeq
\beq
H_V\approx H_b\frac{M_1}{2\sqrt{\Lambda_1}}
\eeq
right after the collision. We see that if the energy density of radiation 
produced on the visible brane is positive, the brane starts to collapse 
rather then expand since $H_b<0$ at the moment of collision.  
This difficulty was discussed in \cite{pyro}. 

\section{Conclusion.}

We have derived a conservation law valid in brane collisions in five 
dimensional anti-deSitter space. Using this conservation law we have analyzed 
a particular situation when a brane sitting at an  orbifold-fixed point
in five-dimensional anti-deSitter space is hit by a bulk brane. We have 
found the energy density of radiation, $\epsilon_{rad}$, 
produced on the orbifold-fixed brane. It is 
expressed through the expansion rate $H_b$ (or, equivalently, kinetic energy)
of the bulk brane at the moment of collision (\ref{result1}). We have found 
that if the tension of the visible brane is positive, the universe 
can start expansion after the collision, while in the case when the 
visible brane has negative tension the universe will start to collapse.

The method described in Section II can be implemented for different background 
metrics. For example, we can develop a procedure, similar to the one 
used in Section III for the bulk metric of the form 
\beq
ds^2=-D(y)dT^2+B^2D^4(y)dy^2+A^2D(y)\l(\sum_{k=1}^3 dx_k^2\r)
\eeq
where $D(y)$ is a given function, as it is in heterotic M-theory motivated 
Ansatz of ekpyrotic/pyrotechnik models.

Since the conservation law (\ref{final}) enables us to relate 
energy density on the visible brane 
after the collision to the energy density of the bulk brane before the
collision, it is useful for the analysis of controversial question of 
behavior of density perturbations in the ekpyrotic models 
\cite{ekpirots,pyro,lyth,brandenberger,ekpirots1}. It is doubtful 
that this question can be resolved within an effective four-dimensional 
theory, since not only the equation of state of matter on the visible brane 
jumps at the moment of collision, but the energy density of matter itself. 
This is quite different from what is usually considered in discussion of
matching the perturbations at the transitions between different 
epochs in the conventional four-dimensional cosmology. 

\section{Acknowledgements.}

I would like to thank the organizers of M-theory Cosmology conference in
Cambridge during which some of this work was done. I am also grateful to 
V.Mukhanov and S.Solodukhin for discussions. This work was supported by 
SFB 375 der Deutsche Forschungsgemeinschaft.

\section{Appendix: Collision of light-like branes (Fig. 2).}

The trajectory of the brane ${\bf b}_1$ in the coordinates
$(U_1, V_1)$ in the region {\bf 1} is given by
equation
\beq
\Sigma_1^+=V_1-\alpha=0
\eeq
where $\alpha$ is a constant. The same trajectory in the coordinates 
$(U_2, V_2)$ in the region {\bf 2} is
\beq
\Sigma_2^-=V_2-\beta=0
\eeq
From the matching condition (\ref{match-H}) we find 
a relation between the coordinates  $U_1$ and $U_2$
\beq
\Lambda_1(\alpha-U_1)=\Lambda_2(\beta-U_2)
\eeq
Differentiating this equation with respect to $u$ gives
\beq
\label{a1}
U_2'=\frac{\Lambda_1}{\Lambda_2}U_1'
\eeq
Using the matching condition (\ref{match-F}) we get also relation 
between $V_1'$ and $V_2'$
\beq
V_2'=V_1'
\eeq
By the same way of reasoning applied to the boundary between regions 
{\bf 2} and {\bf 3} we get relations
\beq
U_3'=U_2';\ \ V_3'=\frac{\Lambda_2}{\Lambda_3}V_2'
\eeq
For the boundary between regions {\bf 3} and {\bf 4} we get
\beq
U_4'=\frac{\Lambda_3}{\Lambda_4}U_3';\ \ 
V_4'=V_3'
\eeq
and for the boundary between {\bf 4} and {\bf 1} we obtain
\beq
\label{a4}
U_1'=U_4'; \ \ 
V_1'=\frac{\Lambda_4}{\Lambda_1}V_4'
\eeq
Combining (\ref{a1})-(\ref{a4}) we find the consistency condition (see 
(\ref{consist}))  
\beq
\label{cons}
\Lambda_1\Lambda_3=\Lambda_2\Lambda_4
\eeq

\end{document}